\documentclass[english]{article}
\usepackage[T1]{fontenc}
\usepackage[latin9]{inputenc}
\usepackage{graphicx}

\makeatletter
\newcommand{\lyxaddress}[1]{
\par {\raggedright #1
\vspace{1.4em}
\noindent\par}
}

\usepackage{a4wide}
\usepackage{hyperref}

\makeatother

\usepackage{babel}
\begin{document}

\title{Diagonal form factors from non-diagonal ones}

\author{Zoltan Bajnok and Chao Wu}

\maketitle

\lyxaddress{\begin{center}
\emph{MTA Lend\"ulet Holographic QFT Group, Wigner Research Centre
for Physics}\\
\emph{Konkoly-Thege Miklós u. 29-33, 1121 Budapest , Hungary}
\par\end{center}}
\begin{abstract}
We prove the asymptotic large volume expression of diagonal form factors
in integrable models by evaluating carefully the diagonal limit of
a non-diagonal form factor in which we send the rapidity of the extra
particle to infinity. 
\end{abstract}

\section{Introduction}

Two dimensional integrable quantum field theories are useful toy models
of statistical and particle physics as they provide many interesting
observables, which can be calculated exactly \cite{Mussardo:2010mgq}.
These models are first solved in infinite volume, where the scattering
matrix \cite{Zamolodchikov:1978xm,Dorey:1996gd}, which connects asymptotic
multiparticle states, are determined together with the form factors
which are the matrix elements of local operators sandwiched between
the same asymptotic states \cite{Smirnov:1992vz}. These form factors
then can be used to build up the correlation function, which define
the theory in the Wightman sense \cite{Babujian:2003sc}. 

In the relevant practical applications, however, quantum field theories
are confined to a finite volume and the calculation of finite size
corrections is unavoidable. Fortunately, all these finite size corrections
can be expressed in terms of the infinite volume characteristics,
such as masses, scattering matrices and form factors \cite{Luscher:1985dn,Luscher:1986pf,Pozsgay:2007kn}.
We can distinguish three domains in the volume according to the nature
of the corrections. The leading finite size corrections are polynomial
in the inverse power of the volume, while the sub-leading corrections
are exponentially volume-suppressed. 

Concerning the finite volume energy spectrum the domain when only
polynomial corrections are kept is called the Bethe-Yang (BY) domain.
We there merely need to take into account the finite volume quantization
of the momenta, which originates from the periodicity requirement
and explicitly includes the scattering phase-shifts \cite{Luscher:1986pf}.
The exponentially small corrections are due to virtual particles traveling
around the world and the domain in which we keep only the leading
exponential correction is called the Luscher domain \cite{Luscher:1985dn}.
In a small volume, when all exponentials contribute the same way,
we have to sum them up leading to a description given by the Thermodynamic
Bethe Ansatz (TBA) \cite{Zamolodchikov:1989cf}. 

The situation for the form factors are not understood at the same
level yet. The BY domain was investigated in \cite{Pozsgay:2007kn,Pozsgay:2007gx}.
It was proven for non-diagonal form factors that all polynomial finite
size effects come only from the finite volume (Kronecker-delta) normalization
of states. The authors also conjectured the BY form of diagonal finite
volume form factors, which they derived for two particle-states. The
leading exponential finite size corrections for generic form factors
are not known, except for the diagonal ones, for which exact conjectures
exist. The LeClair-Mussardo (LM) conjecture expresses the exact finite
volume/temperature one-point functions in terms of infinite volume
diagonal connected form factors, and densities of mirror states determined
by the TBA equation \cite{Leclair:1999ys}. Actually it was shown
in \cite{Pozsgay:2010xd,Pozsgay:2013jua} that the BY form of diagonal
form factors implies the LM formula and vice versa. Using analytical
continuation a 'la \cite{Dorey:1996re} Pozsgay extended the LM formula
for finite volume diagonal matrix elements \cite{Pozsgay:2014gza}.
The aim of the present paper is to prove the conjectured BY form of
diagonal form factors \cite{Saleur:1999hq,Pozsgay:2007gx} from the
already proven non-diagonal BY form factors \cite{Pozsgay:2007kn}
by carefully calculating the diagonal limit, in which we send one
particle's rapidity to infinity. By this way our result also leads
to the prove of the LM formula. Here we focus on theories with one
type of particles. 

The paper is organized such that in the next section we summarize
the know facts about the BY form of diagonal and non-diagonal form
factors. We then in section 3 prove the diagonal conjecture and conclude
in section 4.

\section{The conjecture for diagonal large volume form factors}

In this section we introduce the infinite volume form factors and
their properties and use them later on to describe the finite volume
form factors in the BY domain.

\subsection{Infinite volume form factors }

Infinite volume form factors are the matrix elements of local operators
sandwiched between asymptotic states $\langle\theta'_{1},\dots,\theta'_{m}\vert\mathcal{O}\vert\theta_{n},\dots,\theta_{1}\rangle$.
We use the rapidity $\theta$ to parametrize the momenta as $p=m\sinh\theta$.
The crossing formula 
\begin{eqnarray}
\langle\theta'_{1},\dots,\theta'_{m}\vert\mathcal{O}\vert\theta_{n},\dots,\theta_{1}\rangle & = & \langle\theta'_{1},\dots,\theta'_{m-1}\vert\mathcal{O}\vert\bar{\theta}'_{m}-i\epsilon,\theta_{n},\dots,\theta_{1}\rangle+\\
 &  & \sum_{i=1}^{n}2\pi\delta(\theta_{m}-\theta_{i})\prod_{j=i+1}^{n}S(\theta_{j}-\theta_{i})\langle\theta'_{1},\dots,\theta'_{m-1}\vert\mathcal{O}\vert\theta_{n-1},\dots,\theta_{1}\rangle\nonumber 
\end{eqnarray}
can be used to express every matrix element in terms of the elementary
form factors
\begin{equation}
\langle0\vert\mathcal{O}\vert\theta_{n},\dots,\theta_{1}\rangle=F_{n}(\theta_{n},\dots,\theta_{1})
\end{equation}
where $\bar{\theta}=\theta+i\pi$ denotes the crossed rapidity and
the two particle S-matrix satisfies $S(\theta)=S(i\pi-\theta)=S(-\theta)^{-1}$.
Infinite volume states are normalized to Dirac $\delta$-functions:
as $\langle\theta'\vert\theta\rangle=2\pi\delta(\theta-\theta')$.
The elementary form factor satisfies the permutation and periodicity
axiom 
\begin{eqnarray}
F_{n}(\theta_{1},\theta_{2},\dots,\theta_{i},\theta_{i+1}\dots,\theta_{n}) & = & S(\theta_{i}-\theta_{i+1})F_{n}(\theta_{1},\theta_{2},\dots,\theta_{i+1},\theta_{i}\dots,\theta_{n})\nonumber \\
 & = & F_{n}(\theta_{2},\dots,\theta_{i},\theta_{i+1}\dots,\theta_{n},\theta_{1}-2i\pi)
\end{eqnarray}
together with the kinematical singularity relation
\begin{equation}
-i\mbox{Res}_{\theta'=\theta}F_{n+2}(\theta'+i\pi,\theta,\theta_{1},\dots,\theta_{n})=(1-\prod_{i=1}^{n}S(\theta-\theta_{i}))F_{n}(\theta_{1},\dots,\theta_{n})
\end{equation}
For scalar operators, when properly normalized, the form factor also
satisfies the cluster property
\begin{equation}
\lim_{\Lambda\to\infty}F_{n+m}(\theta_{1}+\Lambda,\dots,\theta_{n}+\Lambda,\theta_{n+1},\dots,\theta_{n+m})=F_{n}(\theta_{1},\dots,\theta_{n})F_{m}(\theta_{n+1},\dots,\theta_{n+m})
\end{equation}
which will be used to analyze the diagonal limit of $\langle\theta,\theta'_{1},\dots,\theta'_{n}\vert\mathcal{O}\vert\theta_{n},\dots,\theta_{1}\rangle$
via $\theta\to\infty$ in finite volume. 

The diagonal form factors $\langle\theta_{1},\dots,\theta{}_{n}\vert\mathcal{O}\vert\theta_{n},\dots,\theta_{1}\rangle$
are singular due to the $\delta(\theta)$ terms coming from the normalization
of the states and also from poles related to the kinematical singularity
axiom. Actually, $F_{2n}(\bar{\theta}_{1}+\epsilon_{1},\dots,\bar{\theta}_{n}+\epsilon_{n},\theta_{n},\dots,\theta_{1})$
is not singular when all $\epsilon_{i}$ go to zero simultaneously,
but depends on the direction of the limit. The \emph{connected} diagonal
form factor is defined as the finite $\epsilon$-independent part:
\begin{equation}
F_{2n}^{c}(\theta_{1},\dots,\theta_{k})=\mbox{Fp}\left(F_{2n}(\bar{\theta}_{1}+\epsilon_{1},\dots,\bar{\theta}_{n}+\epsilon_{n},\theta_{n},\dots,\theta_{1})\right)
\end{equation}
while the \emph{symmetric} evaluation is simply 
\begin{equation}
F_{2n}^{s}(\theta_{1},\dots,\theta_{k})=\lim_{\epsilon\to0}F_{2n}(\bar{\theta}_{1}+\epsilon,\dots,\bar{\theta}_{n}+\epsilon,\theta_{n},\dots,\theta_{1})
\end{equation}
In order to understand the singularity structure of the diagonal limit
we note that the singular part can very nicely be visualized by graphs
\cite{Pozsgay:2007gx}: 
\begin{equation}
F_{2n}(\bar{\theta}_{1}+\epsilon_{1},\dots,\bar{\theta}_{n}+\epsilon_{n},\theta_{n},\dots,\theta_{1})=\sum_{\mathrm{allowed\, graphs}}F(\mathrm{graph)}+O(\epsilon_{i})\label{eq:F_2n=00003Dgraph}
\end{equation}
where an allowed graph is an oriented tree-like (no-loop) graph in
which at each vertex there is at most one outgoing edge. The contribution
of a graph, $F(\mathrm{graph)}$, can be evaluated as follows: points
$(i_{1},\dots,i_{k})$ with no outgoing edges contribute a factor,
$F_{2k}^{c}(\theta_{i_{1}},\dots,\theta_{i_{k}})$, while for each
edge from $i$ to $j$ we associate a factor $\frac{\epsilon_{j}}{\epsilon_{i}}\phi(\theta_{i}-\theta_{j})$,
where $\phi(\theta)=-i\partial_{\theta}\log S(\theta)=-i\frac{S'(\theta)}{S(\theta)}$.
We recall the proof of (\ref{eq:F_2n=00003Dgraph}) from \cite{Pozsgay:2007gx}
as similar argumentations will be used later on. The proof goes in
induction in $n$ and evaluates the residue at $\epsilon_{n}=0$ keeping
all other $\epsilon$s finite. Clearly such singular term can come
only from graphs in which $n$ has only an outgoing edge and no incoming
one. The contributions of such terms are 
\begin{equation}
\frac{1}{\epsilon_{n}}\left(\epsilon_{1}\phi_{1n}+\dots+\epsilon_{n-1}\phi_{n-1n}\right)F_{2n-2}(\bar{\theta}_{1}+\epsilon_{1},\dots,\bar{\theta}_{n-1}+\epsilon_{n-1},\theta_{n-1},\dots,\theta_{1})
\end{equation}
Now comparing this expression to the kinematical singularity axiom
and using the definition of $\phi(\theta)$ together with the properties
of the scattering matrix we can see that they completely agree. The
formula (\ref{eq:F_2n=00003Dgraph}) can be used to define connected
form factors recursively by subtracting the singular terms and taking
the diagonal limit. Observe also that taking all $\epsilon$ to be
the same the lhs. of (\ref{eq:F_2n=00003Dgraph}) becomes the symmetric
form factor, which is expressed by (\ref{eq:F_2n=00003Dgraph}) in
terms of the connected ones. 

In particular, for the 2-particle form factor we have only three graphs:\\

\begin{center}
\includegraphics[width=6cm]{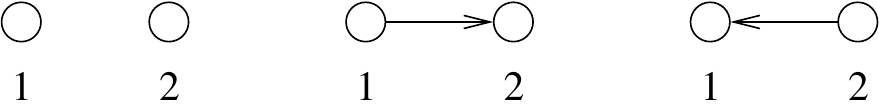}
\par\end{center}

\noindent which give 
\begin{equation}
F_{4}(\bar{\theta}_{1}+\epsilon_{1},\bar{\theta}_{2}+\epsilon_{2},\theta_{2},\theta_{1})=F_{4}^{c}(\theta_{1},\theta_{2})+\frac{\epsilon_{1}}{\epsilon_{2}}\phi_{12}F_{2}^{c}(\theta_{1})+\frac{\epsilon_{2}}{\epsilon_{1}}\phi_{21}F_{2}^{c}(\theta_{2})+O(\epsilon_{i})
\end{equation}
This equation on the one hand can be used to define $F_{4}^{c}(\theta_{1},\theta_{2})$,
once $F_{2}^{c}(\theta)$ has been already defined, and on the other
hand, it connects the symmetric form factor to the connected one:
\begin{equation}
F_{4}^{s}(\theta_{1},\theta_{2})=F_{4}^{c}(\theta_{1},\theta_{2})+\phi_{12}F_{2}^{c}(\theta_{1})+\phi_{21}F_{2}^{c}(\theta_{2})
\end{equation}

\subsection{Finite volume form factors in the BY domain}

\noindent In the BY domain we drop the exponentially suppressed $O(e^{-mL})$
terms and keep only the $O(L^{-1})$ polynomial volume dependence.
The quantization of the momenta is given by the BY equations 
\begin{equation}
Q_{j}\equiv p(\theta_{j})L-i\sum_{k:k\neq j}\log S(\theta_{j}-\theta_{k})=2\pi I_{j}
\end{equation}
An $n$-particle state is labeled by the integers $I_{j}$, which
can be traded for the momenta: $\vert I_{1},\dots,I_{n}\rangle\equiv\vert\theta_{1},\dots,\theta_{n}\rangle_{L}$.
These states are normalized to Kronecker delta functions $\langle I'\vert I\rangle=\prod_{j}\delta_{I_{j}'I_{j}}$.
Since two point functions in finite and infinite volume are equal
up to exponentially small $O(e^{-mL})$ terms the finite and infinite
volume form factors differ only in the normalization of states \cite{Pozsgay:2007kn}.
In particular, this implies the non-diagonal finite volume form factor
formula 
\begin{equation}
\langle\theta_{1}',\dots,\theta_{m}'\vert\mathcal{O}\vert\theta_{n},\dots,\theta_{1}\rangle_{L}=\frac{F_{n+m}(\bar{\theta}_{1}',\dots,\bar{\theta}_{m}',\theta{}_{n},\dots,\theta{}_{1})}{\sqrt{\rho_{n}\rho_{m}'}}+O(e^{-mL})
\end{equation}
where the densities of states are defined through the Bethe Ansatz
equation via 
\begin{equation}
\rho_{n}=\det\vert Q_{ij}\vert\quad;\qquad Q_{ij}=\partial_{i}Q_{j}\equiv\frac{\partial Q_{j}}{\partial\theta_{i}}
\end{equation}
 The conjectured formula for diagonal form factors takes the form
\cite{Saleur:1999hq}:
\begin{equation}
\langle\theta_{1},\dots,\theta_{n}\vert\mathcal{O}\vert\theta_{n},\dots,\theta_{1}\rangle_{L}=\frac{\sum_{\alpha\cup\bar{\alpha}}F_{\alpha}^{c}\rho_{\bar{\alpha}}}{\rho_{n}}+O(e^{-mL})
\end{equation}
where the index set $I=\{1,\dots,n\}$ is split in all possible ways
$I=\alpha\cup\bar{\alpha}$, $F_{\alpha}^{c}=F_{2k}^{c}(\theta_{\alpha_{1}},\dots,\theta_{\alpha_{k}})$
with $\vert\alpha\vert=k$ and $\rho_{\bar{\alpha}}$ is the shorthand
for $\rho_{n-k}(\theta_{\bar{\alpha}_{1}},\dots,\theta_{\bar{\alpha}_{n-k}})$,
which denotes the sub-determinant of the matrix, $Q_{ij}$, with indices
only from $\bar{\alpha}$. There is an analogous expression in terms
of the symmetric form factors \cite{Pozsgay:2007gx} 
\begin{equation}
\langle\theta_{1},\dots,\theta_{n}\vert\mathcal{O}\vert\theta_{n},\dots,\theta_{1}\rangle_{L}=\frac{\sum_{\alpha\cup\bar{\alpha}}F_{\alpha}^{s}\rho_{\bar{\alpha}}^{s}}{\rho_{n}}+O(e^{-mL})
\end{equation}
where now $\rho_{\alpha}^{s}$ is the density of states corresponding
to the variables with labels in $\alpha$. The equivalence of the
two formulas was shown in \cite{Pozsgay:2007gx}. Let us note that
for $L=0$ the sum reduces to one single term $\sum_{\alpha\cup\bar{\alpha}}F_{\alpha}^{s}\rho_{\bar{\alpha}}^{s}\to F_{n}^{s}$
as all other $\rho^{s}$ factor vanish. 

Let us spell out the details for two particles. The diagonal finite
volume form factor is 
\begin{equation}
\langle\theta_{1},\theta_{2}\vert\mathcal{O}\vert\theta_{2},\theta_{1}\rangle_{L}=\frac{F_{4}^{c}(\theta_{1},\theta_{2})+\rho_{1}(\theta_{1})F_{2}^{c}(\theta_{2})+\rho_{1}(\theta_{2})F_{2}^{c}(\theta_{1})+\rho_{2}(\theta_{1},\theta_{2})F_{0}}{\rho_{2}(\theta_{1},\theta_{2})}+O(e^{-mL})
\end{equation}
where 
\[
\rho_{2}(\theta_{1},\theta_{2})=\left|\begin{array}{cc}
E_{1}L+\phi_{12} & -\phi_{12}\\
-\phi_{21} & E_{2}L+\phi_{21}
\end{array}\right|\quad;\qquad\rho_{1}(\theta_{i})=E_{i}L+\phi_{i3-i}
\]
The analogue formula with the symmetric evaluation reads as
\begin{equation}
\langle\theta_{1},\theta_{2}\vert\mathcal{O}\vert\theta_{2},\theta_{1}\rangle_{L}=\frac{F_{4}^{s}(\theta_{1},\theta_{2})+\rho_{1}^{s}(\theta_{1})F_{2}^{s}(\theta_{2})+\rho_{1}^{s}(\theta_{2})F_{2}^{s}(\theta_{1})+\rho_{2}^{s}(\theta_{1},\theta_{2})F_{0}^{s}}{\rho_{2}(\theta_{1},\theta_{2})}+O(e^{-mL})
\end{equation}
where 
\[
\rho_{2}^{s}(\theta_{1},\theta_{2})=\rho_{2}(\theta_{1},\theta_{2})\quad;\qquad\rho_{1}^{s}(\theta_{i})=E_{i}L
\]

\section{The proof for diagonal large volume form factors}

The idea of the proof follows from the large $\theta$ behaviour of
the scattering matrix, namely $S(\theta)\to1$, for $\theta\to\infty$.
This also lies behind the cluster property of the form factors. Thus
by taking the non-diagonal form factor $\langle\theta,\theta_{1}',\dots,\theta_{n}'\vert\mathcal{O}\vert\theta{}_{n},\dots,\theta{}_{1}\rangle_{L}$
and sending $\theta\to\infty$, the extra particle decouples and we
can approach the diagonal form factor. This can be achieved by choosing
the same quantization numbers for both the $\theta_{j}$ and $\theta'_{j}$
particles: 
\begin{equation}
Q{}_{j}'\equiv p(\theta_{j}')L-i\sum_{k:k\neq j}\log S(\theta{}_{j}'-\theta{}_{k}')-i\log S(\theta{}_{j}'-\theta)=2\pi I_{j}
\end{equation}
Indeed, by sending (the quantization number of) $\theta$ to infinity
the BA equations, $Q_{j}'$, reduce to the Bethe Ansatz equations,
$Q_{j}$. This means that in the limit considered $\theta{}_{i}'=\theta_{i}+\epsilon_{i}$
and $\epsilon_{i}$ goes to zero. In principle, $\epsilon_{i}$ depends
on $\{\theta_{i}\}$ and on the way how $\theta$ goes to infinity. 

For finite $\theta$, the form factor is non-diagonal and we can use
\begin{equation}
\langle\theta,\theta_{1}',\dots,\theta_{n}'\vert\mathcal{O}\vert\theta{}_{n},\dots,\theta{}_{1}\rangle_{L}=\frac{F_{2n+1}(\bar{\theta},\bar{\theta}_{1}',\dots,\bar{\theta}_{n}',\theta_{n},\dots,\theta_{1})}{\sqrt{\rho'_{n+1}\rho_{n}}}+O(e^{-mL})
\end{equation}
The numerator is a finite quantity for any $\theta$ and has a finite
$\theta\to\infty$ limit accordingly. We can see in the limit that
$\rho'_{n+1}(\theta,\theta_{1}',\dots,\theta_{n}')$ goes to $\rho_{1}(\theta)\rho_{n}(\theta_{1},\dots,\theta_{n})$.
Similarly for the form factors, the cluster property guaranties that
$F_{2n+1}(\bar{\theta},\bar{\theta}_{1}',\dots,\bar{\theta}_{n}',\theta_{n},\dots,\theta_{1})$
will be factorized as $F_{2n}(\bar{\theta}_{1}',\dots,\bar{\theta}_{n}',\theta_{n},\dots,\theta_{1})F_{1}(\bar{\theta})$,
where additionally $\theta_{i}'\to\theta_{i}$. Actually the expression
depends in the direction we take the limit in which all $\epsilon_{i}$
go to zero and our main task is to calculate this limit explicitly.
Fortunately, the direction is dictated by the difference of the BA
equations: 
\begin{equation}
Q'_{j}-Q_{j}=E_{j}L\epsilon_{j}+\sum_{k:k\neq j}\phi_{jk}(\epsilon_{j}-\epsilon_{k})-\delta_{j}=\sum_{k}Q_{jk}\epsilon_{k}-\delta_{j}=0
\end{equation}
where we have streamlined the notations: 
\begin{equation}
E_{j}=\partial_{j}p(\theta_{j})\quad;\qquad\phi_{jk}=\phi(\theta_{j}-\theta_{k})=-i\partial_{j}\log S(\theta_{j}-\theta_{k})\quad;\qquad\delta_{j}=i\log S(\theta_{j}-\theta)
\end{equation}
Clearly $\delta_{j}$s are small and so are the $\epsilon_{j}$s.
In the following we analyze the $\epsilon$ and $\delta$ dependence
of the form factor $F_{2n+1}(\bar{\theta},\bar{\theta}_{1}',\dots,\bar{\theta}_{n}',\theta_{n},\dots,\theta_{1})$.
Similarly to the diagonal limit of form factors we can describe the
$\delta$ and $\epsilon$ dependence by graphs. We claim that

\begin{equation}
F_{2n+1}(\bar{\theta},\bar{\theta}_{1}',\dots,\bar{\theta}_{k}',\theta_{k},\dots,\theta_{1})=\sum_{\mathrm{allowed\, graphs,colorings}}F(\mathrm{graph)}+O(\epsilon_{i},\delta)\label{eq:F_2n+1=00003Dgraph}
\end{equation}
where, additionally to the previous graphs in (\ref{eq:F_2n=00003Dgraph}),
we should allow the coloring of those vertices, which do not have
any outgoing edge, i.e. they can come either in black or in white.
For each black dot with label $i$ we associate a factor $\frac{\delta_{i}}{\epsilon_{i}}$.
Note that in the $\theta\to\infty$ limit we will have an overall
$F_{1}(\bar{\theta})$ factor, which we factor out. 

Let us see how it works for $n=1$: The single dot can be either black
or white: \\

\begin{center}
\includegraphics[width=2cm]{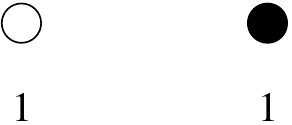}
\par\end{center}

\noindent thus the two contributions are 
\begin{equation}
F_{3}(\bar{\theta},\bar{\theta}_{1}',\theta_{1})F_{1}(\bar{\theta})^{-1}=\frac{\delta_{1}}{\epsilon_{1}}+F_{2}^{c}(\theta_{1})+\dots
\end{equation}
where ellipses represents terms vanishing in the $\delta,\epsilon\to0$
limit. Let us show that $F_{2}^{c}(\theta_{1})$ is non singular,
i.e. the singularity of the lhs. is exactly $\frac{\delta_{1}}{\epsilon_{1}}.$
The kinematical residue equation tells us that 
\begin{equation}
F_{3}(\bar{\theta},\bar{\theta}_{1}',\theta_{1})=\frac{i}{\epsilon_{1}}(1-S(\theta_{1}'-\theta+i\pi))F_{1}(\bar{\theta})+O(1)=\frac{\delta_{1}}{\epsilon_{1}}F_{1}(\bar{\theta})+O(1)
\end{equation}
Thus, once the singularity is subtracted, we can safely take the $\epsilon_{1}\to0$
and the $\delta\to0$ limits leading to 
\begin{equation}
\lim_{\delta,\epsilon_{1}\to0}(F_{3}(\bar{\theta},\bar{\theta}_{1}',\theta_{1})-\frac{\delta_{1}}{\epsilon_{1}}F_{1}(\bar{\theta}))=F_{2}^{c}(\theta_{1})F_{1}(\bar{\theta})
\end{equation}
where we used the cluster property of form factors and the fact that
the two particle diagonal connected form factor is non-singular. 

Now we adapt the proof in the induction step in (\ref{eq:F_2n=00003Dgraph})
by noticing that the $\epsilon_{n}^{-1}$ singularity can come either
from terms with only one outgoing edge or from being black. Thus the
residue is 
\begin{equation}
\frac{1}{\epsilon_{n}}\left(\delta_{n}+\epsilon_{1}\phi_{1n}+\dots+\epsilon_{n-1}\phi_{n-1n}\right)F_{2n-1}(\bar{\theta},\bar{\theta}_{1}+\epsilon_{1},\dots,\bar{\theta}_{n-1}+\epsilon_{n-1},\theta_{n-1},\dots,\theta_{1})
\end{equation}
 Let us calculate the analogous term from the kinematical residue
axiom:
\begin{eqnarray}
F_{2n+1}(\bar{\theta},\bar{\theta}_{1}',\dots,\bar{\theta}_{n}',\theta_{n},\dots,\theta_{1}) & \to & \frac{i}{\epsilon_{n}}\left(1-\frac{S(\theta_{n}'-\theta_{n-1})\dots S(\theta_{n}'-\theta_{1})}{S(\theta_{n}'-\theta_{n-1}')\dots S(\theta_{n}'-\theta_{1}')}\frac{1}{S(\theta_{n}'-\theta)}\right)\times\nonumber \\
 &  & \,\,\,\,\,\,\, F_{2n-1}(\bar{\theta},\bar{\theta}_{1}',\dots,\bar{\theta}_{n-1}',\theta_{n-1},\dots,\theta_{1})
\end{eqnarray}
The bracket can be expanded as 
\begin{equation}
\left(\right)=-i(\delta_{n}+\phi_{nn-1}\epsilon_{n-1}+\dots+\phi_{n1}\epsilon_{1})\label{eq:bracket}
\end{equation}
which completes the induction.

In particular, for two particles we have the following diagrams:\\

\includegraphics[width=7cm]{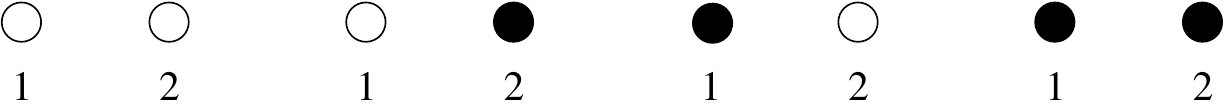}~~~~~~~~~~~~\includegraphics[width=7cm]{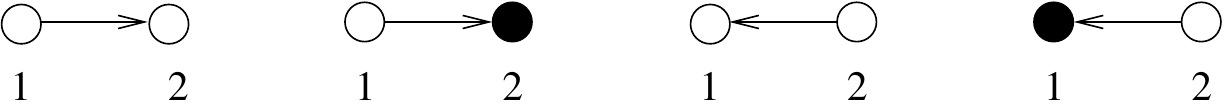}

\noindent which lead to the formula 
\begin{eqnarray}
F_{5}(\bar{\theta},\bar{\theta}_{1}',\bar{\theta}_{2}',\theta_{2},\theta_{1})F_{1}^{-1} & = & F_{4}^{c}(\theta_{1},\theta_{2})+\frac{\epsilon_{2}}{\epsilon_{1}}\phi_{21}\frac{\delta_{2}}{\epsilon_{2}}+\frac{\epsilon_{1}}{\epsilon_{2}}\phi_{12}\frac{\delta_{1}}{\epsilon_{1}}+\frac{\delta_{1}}{\epsilon_{1}}\frac{\delta_{2}}{\epsilon_{2}}\\
 &  & +\frac{\epsilon_{1}}{\epsilon_{2}}\phi_{12}F_{2}^{c}(\theta_{1})+\frac{\delta_{2}}{\epsilon_{2}}F_{2}^{c}(\theta_{1})+\frac{\epsilon_{2}}{\epsilon_{1}}\phi_{21}F_{2}^{c}(\theta_{2})+\frac{\delta_{1}}{\epsilon_{1}}F_{2}^{c}(\theta_{2})\nonumber 
\end{eqnarray}
It is interesting to check the coefficient of $F_{2}^{c}(\theta_{1}):$
\begin{equation}
\frac{\epsilon_{1}\phi_{12}+\delta_{2}}{\epsilon_{2}}=E_{2}L+\phi_{21}=\rho_{1}(\theta_{2})
\end{equation}
where we used the BA equations. Similarly 
\begin{equation}
\frac{\delta_{1}}{\epsilon_{1}}\frac{\delta_{2}}{\epsilon_{2}}+\frac{\epsilon_{1}}{\epsilon_{2}}\phi_{12}\frac{\delta_{1}}{\epsilon_{1}}+\frac{\epsilon_{2}}{\epsilon_{1}}\phi_{21}\frac{\delta_{2}}{\epsilon_{2}}=\rho_{2}(\theta_{1},\theta_{2})
\end{equation}
 which leads to the seeked for formula for $n=2$:
\begin{equation}
F_{5}(\bar{\theta},\bar{\theta}_{1}',\bar{\theta}_{2}',\theta_{2},\theta_{1})F_{1}^{-1}=F_{4}^{c}(\theta_{1},\theta_{2})+\rho_{1}(\theta_{2})F_{2}^{c}(\theta_{1})+\rho_{1}(\theta_{1})F_{2}^{c}(\theta_{2})+\rho_{2}(\theta_{1},\theta_{2})
\end{equation}

In the following we prove the form of the diagonal form factors in
the general case by induction. First we notice that once we use the
BA equations to express $\delta_{i}$ in terms of $\epsilon_{k}$
then all denominators of $\epsilon s$ disappear. Focus on $\epsilon_{n}^{-1}$
and observe that 
\begin{equation}
\delta_{n}+\epsilon_{1}\phi_{1n}+\dots+\epsilon_{n-1}\phi_{n-1n}=\epsilon_{n}\left(E_{n}L+\phi_{n-1n}+\dots+\phi_{1n}\right)
\end{equation}
This implies that the diagonal finite volume form factor is a polynomial
in $L$ and linear in each $E_{k}L$. We first check the $L=0$ piece
and then calculate the derivative wrt. $E_{n}L$ as the full expression
is symmetric in all variables. Note that the would be singular term
in $\epsilon_{n}$ at $L=0$: 
\begin{equation}
\frac{1}{\epsilon_{n}}\epsilon_{n}\left(E_{n}L+\phi_{n-1n}+\dots+\phi_{1n}\right)\vert_{L=0}=\frac{1}{\epsilon}\left(\epsilon\phi_{n-1n}+\dots+\epsilon\phi_{1n}\right)
\end{equation}
is exactly the same we would obtain if we had calculated the diagonal
limit of the form factor in the symmetric evaluation, i.e. for $L=0$
we obtain the symmetric $n$-particle form factor. We now check the
linear term in $E_{n}L$. In doing so we differentiate the expression
(\ref{eq:F_2n+1=00003Dgraph}) wrt. $E_{n}L$:
\begin{equation}
\partial_{E_{n}L}F_{2n+1}(\bar{\theta},\bar{\theta}_{1}',\dots,\bar{\theta}_{n}',\theta_{n},\dots,\theta_{1})=F_{2n-1}(\bar{\theta},\bar{\theta}_{1}',\dots,\bar{\theta}_{n-1}',\theta_{n-1},\dots,\theta_{1})
\end{equation}
since the term $E_{n}L$ can come only through the singularity at
$\epsilon_{n}=0$. Note that on the rhs. $\theta_{k}$ satisfies the
original BA and not the one where $\theta_{n}$ is missing. Let us
now take a look of the expression we would like to prove:
\begin{equation}
F_{2n+1}(\bar{\theta},\bar{\theta}_{1}',\dots,\bar{\theta}_{n}',\theta_{n},\dots,\theta_{1})F_{1}^{-1}=\sum_{\alpha\cup\bar{\alpha}}F_{\alpha}^{c}\rho_{\bar{\alpha}}=\sum_{\alpha\cup\bar{\alpha}}F_{\alpha}^{s}\rho_{\bar{\alpha}}^{s}
\end{equation}
Clearly the rhs. is also a polynomial in $L$, which is linear in
each $E_{k}L$. To finish the proof, we note that the $L=0$ constant
part of the rhs. is the symmetric form factor. Using that $\partial_{E_{n}L}\rho_{\alpha}=\rho_{\alpha\setminus\{n\}}$
if $n\in\alpha$ and 0 otherwise we can see that 
\begin{equation}
\partial_{E_{n}L}\sum_{\alpha\cup\bar{\alpha}=\{1,\dots,n\}}F_{\alpha}^{c}\rho_{\bar{\alpha}}=\sum_{\beta\cup\bar{\beta}=\{1,\dots,n-1\}}F_{\beta}^{c}\rho_{\bar{\beta}}=F_{2n-1}(\bar{\theta},\bar{\theta}_{1}',\dots,\bar{\theta}_{n-1}',\theta_{n-1},\dots,\theta_{1})F_{1}^{-1}
\end{equation}
 by the induction hypothesis, which completes the proof.

\section{Conclusion}

In this paper we proved the large volume expression for the diagonal
form factors by taking carefully the limit of a nondiagonal form factor.
Our result completes the proof of the LM formula, which describes
exactly the one-point function in a finite volume. 

Diagonal finite volume form factors are relevant in the AdS/CFT correspondence
as they conjectured to describe the Heavy-Heavy-Light (HHL) type three
point functions of the maximally supersymmetric 4D gauge theory \cite{Bajnok:2014sza}.
This conjecture was first proved at weak coupling \cite{Hollo:2015cda}
then at strong coupling \cite{Bajnok:2016xxu}, finally for all couplings
in \cite{Jiang:2015bvm,Jiang:2016dsr}. We profited from all of these
proofs in the present paper. 

There is a natural extension of our results for diagonal form factors
in non-diagonal theories. Clearly the same idea of adding one more
particle and sending its rapidity to infinite can be applied there
too and we have ongoing research into this direction.

\section*{Acknowledgments}

We thank Yunfeng Jiang the enlightening discussions and ZB thanks
the hospitality of the mathematical research institute MATRIX in Australia
where the paper was finalized. The work was supported by a Lendület
and by the NKFIH 116505 Grant.


\begin{thebibliography}{10}

\bibitem{Mussardo:2010mgq}
Giuseppe Mussardo.
\newblock {\em {Statistical field theory}}.
\newblock Oxford Univ. Press, New York, NY, 2010.

\bibitem{Zamolodchikov:1978xm}
Alexander~B. Zamolodchikov and Alexei~B. Zamolodchikov.
\newblock {Factorized s Matrices in Two-Dimensions as the Exact Solutions of
  Certain Relativistic Quantum Field Models}.
\newblock {\em Annals Phys.}, 120:253--291, 1979.

\bibitem{Dorey:1996gd}
P.~Dorey.
\newblock {Exact S matrices}.
\newblock hep-th/9810026.

\bibitem{Smirnov:1992vz}
F.A. Smirnov.
\newblock {Form-factors in completely integrable models of quantum field
  theory}.
\newblock {\em Adv.Ser.Math.Phys.}, 14:1--208, 1992.

\bibitem{Babujian:2003sc}
H.~Babujian and M.~Karowski.
\newblock {Towards the construction of Wightman functions of integrable quantum
  field theories}.
\newblock {\em Int. J. Mod. Phys.}, A19S2:34--49, 2004.

\bibitem{Luscher:1985dn}
M.~Luscher.
\newblock {Volume Dependence of the Energy Spectrum in Massive Quantum Field
  Theories. 1. Stable Particle States}.
\newblock {\em Commun. Math. Phys.}, 104:177, 1986.

\bibitem{Luscher:1986pf}
M.~Luscher.
\newblock {Volume Dependence of the Energy Spectrum in Massive Quantum Field
  Theories. 2. Scattering States}.
\newblock {\em Commun. Math. Phys.}, 105:153--188, 1986.

\bibitem{Pozsgay:2007kn}
B.~Pozsgay and G.~Takacs.
\newblock {Form-factors in finite volume I: Form-factor bootstrap and truncated
  conformal space}.
\newblock {\em Nucl. Phys.}, B788:167--208, 2008.

\bibitem{Zamolodchikov:1989cf}
A.~B. Zamolodchikov.
\newblock {Thermodynamic Bethe Ansatz in Relativistic Models. Scaling Three
  State Potts and Lee-yang Models}.
\newblock {\em Nucl. Phys.}, B342:695--720, 1990.

\bibitem{Pozsgay:2007gx}
B.~Pozsgay and G.~Takacs.
\newblock {Form factors in finite volume. II. Disconnected terms and finite
  temperature correlators}.
\newblock {\em Nucl. Phys.}, B788:209--251, 2008.

\bibitem{Leclair:1999ys}
A.~Leclair and G.~Mussardo.
\newblock {Finite temperature correlation functions in integrable QFT}.
\newblock {\em Nucl. Phys.}, B552:624--642, 1999.

\bibitem{Pozsgay:2010xd}
Balazs Pozsgay.
\newblock {Mean values of local operators in highly excited Bethe states}.
\newblock {\em J. Stat. Mech.}, 1101:P01011, 2011.

\bibitem{Pozsgay:2013jua}
Balazs Pozsgay.
\newblock {Form factor approach to diagonal finite volume matrix elements in
  Integrable QFT}.
\newblock {\em JHEP}, 07:157, 2013.

\bibitem{Dorey:1996re}
Patrick Dorey and Roberto Tateo.
\newblock {Excited states by analytic continuation of TBA equations}.
\newblock {\em Nucl. Phys.}, B482:639--659, 1996.

\bibitem{Pozsgay:2014gza}
B.~Pozsgay, I.~M. Szecsenyi, and G.~Takacs.
\newblock {Exact finite volume expectation values of local operators in excited
  states}.
\newblock {\em JHEP}, 04:023, 2015.

\bibitem{Saleur:1999hq}
H.~Saleur.
\newblock {A Comment on finite temperature correlations in integrable QFT}.
\newblock {\em Nucl. Phys.}, B567:602--610, 2000.

\bibitem{Bajnok:2014sza}
Zoltan Bajnok, Romuald~A. Janik, and Andrzej Wereszczynski.
\newblock {HHL correlators, orbit averaging and form factors}.
\newblock {\em JHEP}, 09:050, 2014.

\bibitem{Hollo:2015cda}
Laszlo Hollo, Yunfeng Jiang, and Andrei Petrovskii.
\newblock {Diagonal Form Factors and Heavy-Heavy-Light Three-Point Functions at
  Weak Coupling}.
\newblock {\em JHEP}, 09:125, 2015.

\bibitem{Bajnok:2016xxu}
Zoltan Bajnok and Romuald~A. Janik.
\newblock {Classical limit of diagonal form factors and HHL correlators}.
\newblock {\em JHEP}, 01:063, 2017.

\bibitem{Jiang:2015bvm}
Yunfeng Jiang and Andrei Petrovskii.
\newblock {Diagonal form factors and hexagon form factors}.
\newblock {\em JHEP}, 07:120, 2016.

\bibitem{Jiang:2016dsr}
Yunfeng Jiang.
\newblock {Diagonal Form Factors and Hexagon Form Factors II. Non-BPS Light
  Operator}.
\newblock {\em JHEP}, 01:021, 2017.

\end{thebibliography}
\end{document}